\documentclass[aps,prd,twocolumn,superscriptaddress,preprintnumbers,reprint,nofootinbib,showpacs]{revtex4}
\usepackage{graphicx}
\usepackage{epsfig}
\usepackage{amssymb}
\begin{document}

\title{Polyakov loop in 2+1 flavor QCD}

\author{A. Bazavov} 
\affiliation{
Physics Department, Brookhaven National Laboratory, Upton, NY 11973, USA
}
\author{P. Petreczky}
\affiliation{
Physics Department, Brookhaven National Laboratory, Upton, NY 11973, USA
}

\begin{abstract}
We study the temperature dependence of the renormalized Polyakov loop
in 2+1 flavor QCD for temperatures $T<210$~MeV. We extend previous 
calculations by the HotQCD collaboration using the highly improved
staggered quark action and perform a
continuum extrapolation of the renormalized Polyakov loop. We compare
the lattice results with the prediction of non-interacting static-light
hadron resonance gas, which describes the temperature dependence of
the renormalized Polyakov loop up to $T<140$ MeV but fails above that
temperature. Furthermore, we discuss the temperature
dependence of the light and strange quark condensates.
\end{abstract}

\maketitle

\section{Introduction}

At high temperature strongly interacting matter undergoes
a transition to a new state characterized by deconfinement
and color screening (see e.g. Refs. \cite{Petreczky:2012rq,Philipsen:2012nu} for recent reviews).
The Polyakov loop is an order parameter for deconfinement
phase transition in $SU(N)$ gauge theories. After
proper renormalization it is related to the free energy
of a static quark $F_Q$ \cite{Kaczmarek:2002mc,Digal:2003jc}.
More precisely, it can be defined through the difference in the free energy
of a system containing a static quark anti-quark 
($Q\bar Q$) pair at infinite separation and a system without static charges
at the same temperature $F_{\infty}(T)$, i.e. 
$L_{ren}(T)=\exp(-F_{\infty}(T)/2T)=\exp(-F_Q/T)$ \cite{Kaczmarek:2002mc}.
In the confined phase $F_{\infty}=\infty$ since the static $Q$ and $\bar Q$ cannot
be separated to infinite distance. Consequently, the Polyakov loop which transforms
non-trivially under the center of the gauge group is zero.
In the deconfined phase the static quark and anti-quark could
be separated to infinite distance due to color screening
which means breaking of the center symmetry $Z(N)$ \cite{McLerran:1981pb,Kuti:1980gh}.
Dynamical quarks explicitly break the $Z(N)$ symmetry of the partition
function and $F_{\infty}$ is finite since the static $Q$ and $\bar Q$
can be now separated to infinite distance by creating a dynamical quark anti-quark pair from
the vacuum, a phenomenon often called string breaking.
In 2+1 flavor QCD  
with the quark masses realized in nature
there is no phase transition related to deconfinement.
Moreover, the renormalized Polyakov loop cannot be
related to the singular part of the free energy density
\cite{Bazavov:2009zn,Bazavov:2011nk}. However, 
the renormalized Polyakov loop is sensitive to
the color screening in hot QCD medium, at high
temperature it is closely related to the 
Debye screening mass (see e.g. \cite{Petreczky:2005bd}).
In the opposite limit of very low temperatures $F_Q$ is related
to the binding energy of a static-light meson (see e.g. \cite{Digal:2001iu}).
Thus, the renormalized Polyakov loop is a good probe of the hot
strongly interacting medium.

On the lattice with temporal extent $N_{\tau}$ the renormalized Polyakov loop is calculated according
to the following formula:
\begin{equation}
L_{ren}(T) = \exp(-c N_{\tau}/2) \left \langle \frac{1}{3} {\rm Tr} \prod_{x_0=1}^{N_\tau} U_0(x_0,\vec{x}) \right \rangle,
\label{Ldef}
\end{equation}
where $U_0(x_0,\vec{x})$ is the gauge link variable in the time direction and $c$ is the lattice spacing
dependent normalization constant that ensures that the static potential calculated on the lattice
has a certain value at a chosen distance \cite{Aoki:2006br}.
In the recent past the renormalized Polyakov loop has been
calculated on the lattice in 2+1 flavor QCD with
physical quark masses using the improved staggered fermion
formulation \cite{Aoki:2006br,Cheng:2007jq,Aoki:2009sc,Bazavov:2009zn,Borsanyi:2010bp,Bazavov:2011nk}.
Furthermore, using the stout improved staggered action continuum
results for the renormalized Polyakov loop
have been presented \cite{Borsanyi:2010bp}.
The aim of this paper is to study the renormalized Polyakov loop
in the low temperature and transition regions and to perform
an independent continuum extrapolation using the highly improved
staggered quark (HISQ) action \cite{Follana:2006rc}.
While the temperature dependence of the Polyakov loop in QCD at high temperatures
is very similar to its temperature dependence in pure gauge theory, this is not
the case for the low temperature and the transition regions.
To understand at which
temperature color screening effects  
set in, it is important to clarify to what extent the temperature
dependence of the Polyakov loop can be understood in terms of hadrons.
As mentioned above, at very low temperatures the dominant contribution to 
$F_Q$ is given by the lowest static-light state. As the temperature increases, more
massive states will contribute as well and also the interactions of 
static-light hadrons with the medium will become more important.
For the description of the bulk thermodynamic quantities it turns out that interactions
between hadrons can be taken into account by adding the contribution of hadronic resonances.
It is reasonable to assume that the effects of interactions of static-light hadrons
with the hadrons in the medium can be accounted for by adding excited (resonance) states.
Therefore, we calculate the renormalized
Polyakov loop in the approximation of non-interacting gas of static-light
hadrons and hadronic resonances as has been suggested recently \cite{Megias:2012kb}.
Contrary to Ref. \cite{Megias:2012kb} (see also \cite{RuizArriola:2012wd,Megias:2012hk}),
where the experimental spectrum of heavy-light(strange) hadrons was used together with
different model considerations, our analysis 
is largely based on the lattice QCD calculations of the spectrum of static-light and static-strange
hadrons \cite{Michael:2010aa,Wagner:2011fs}. 
We also consider different quark model analyses of the heavy-light(strange) hadron spectrum,
compare them with each other and the available lattice calculations, and use them to estimate
the contribution of higher lying excited states to $L_{ren}$.

The rest of the paper is organized as follows.
In section II we present our numerical results for the renormalized Polyakov loop.
The temperature dependence of the quark condensates is also discussed there.
In section III we discuss the spectrum of static-light hadrons and the calculation
of the Polyakov loop using the hadron resonance gas approximation for static-light
hadrons.  Finally, section IV contains our conclusions. 
 
\section{Numerical results}
\begin{figure*}
\includegraphics[width=8.6cm]{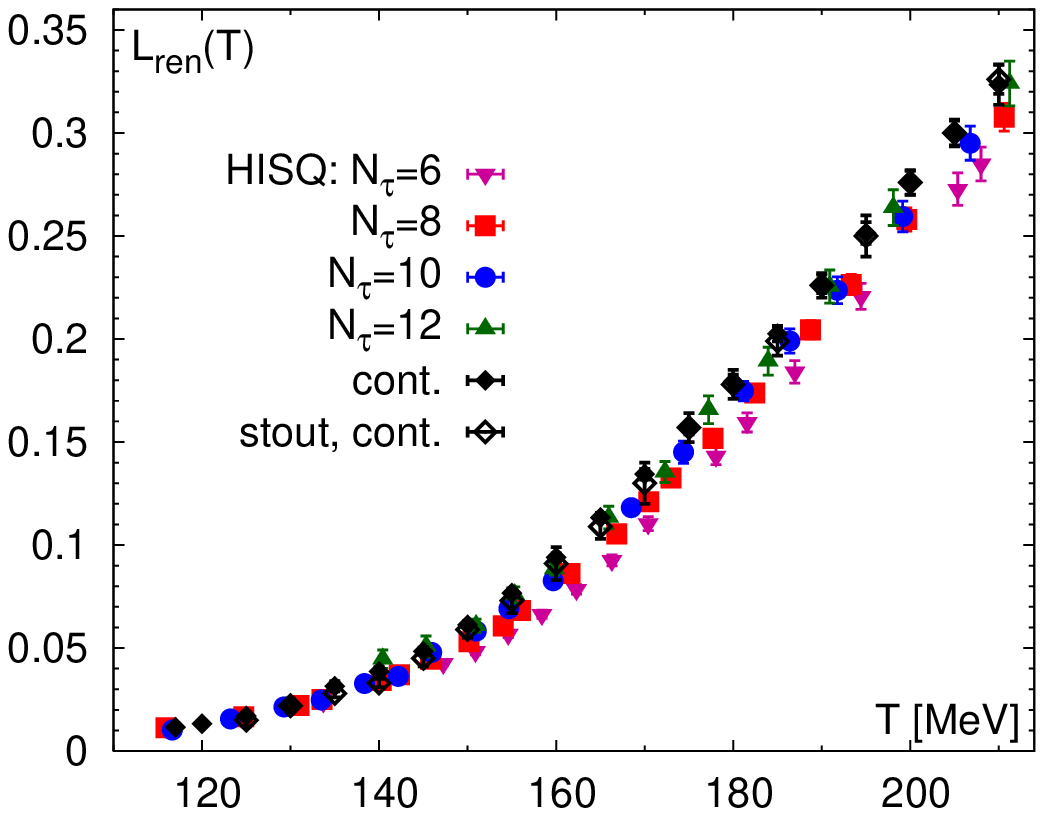}
\includegraphics[width=8.6cm]{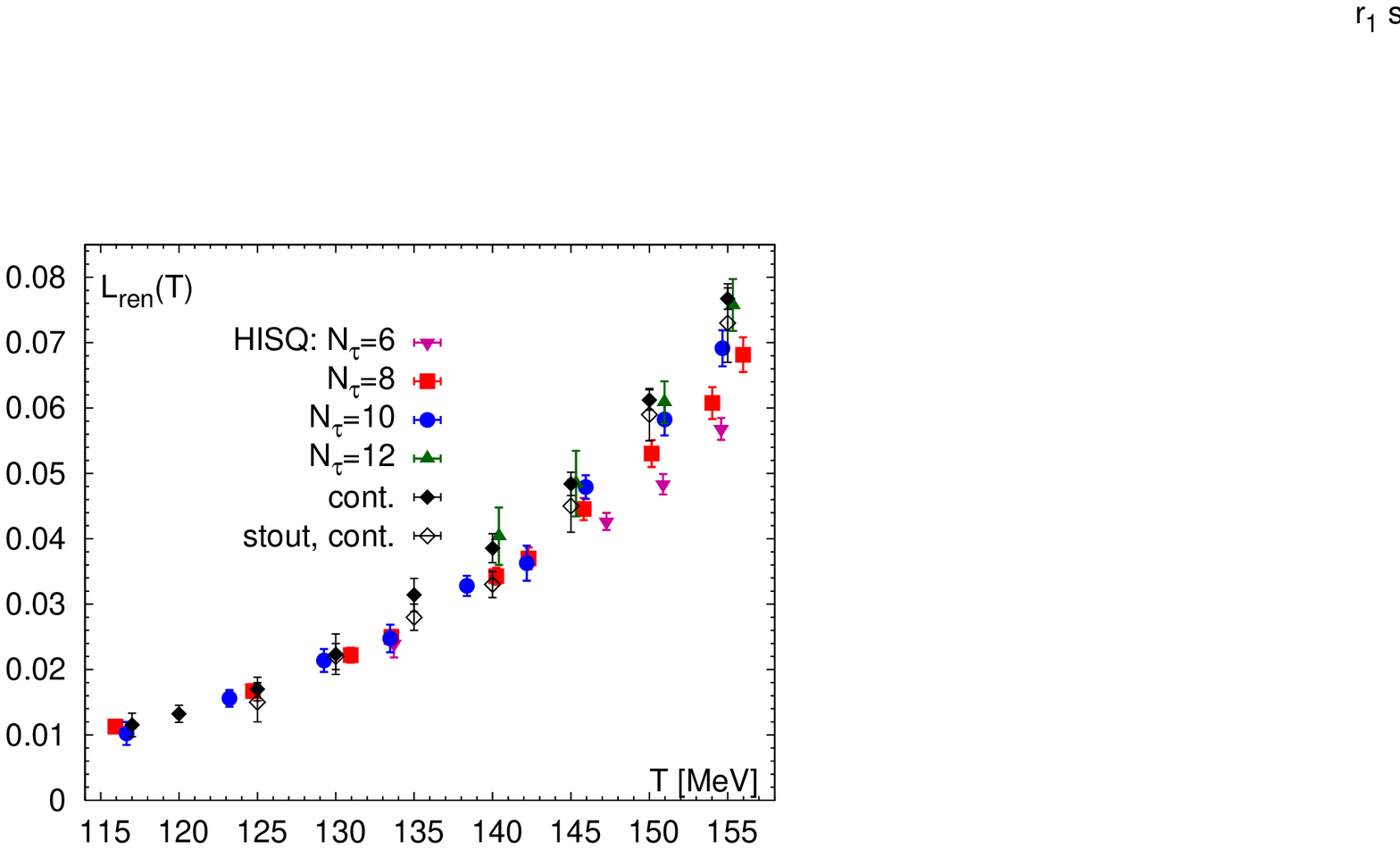}
\caption{The renormalized Polyakov loop calculated with the HISQ and stout action.
The HISQ results for $N_{\tau}=6,~8$ and $12$ are from Ref. \cite{Bazavov:2011nk}.
The continuum stout data are from Ref. \cite{Borsanyi:2010bp}. The filled
diamonds correspond to the continuum extrapolation for the HISQ action.
The right panel shows the closeup of the Polyakov loop in the low temperature region.}
\label{fig:lren}
\end{figure*}

The chiral and deconfining aspects of the QCD transition have been studied  by
the HotQCD collaboration using lattices with temporal extent $N_{\tau}=6,~8$, and $12$
and a combination of the tree-level improved gauge action and the HISQ action in the quark sector 
\cite{Bazavov:2011nk}. This combination of the gauge action and quark action was referred
to as the HISQ/tree action in Ref. \cite{Bazavov:2011nk} but here we refer to it
as the HISQ action for simplicity. 
For reliable continuum extrapolations we need at least three lattice spacings.
Therefore, we performed calculations using $40^3 \times 10$ lattices,
using as in the earlier work the
rational hybrid Monte-Carlo algorithm \cite{Clark:2004cp}.
The algorithmic details of dynamical HISQ simulations can be found in Ref. \cite{Bazavov:2010ru}.
As in Ref. \cite{Bazavov:2011nk}, calculations are performed for the physical value of
the strange quark mass $m_s$ and light quark masses $m_l=m_s/20$. This light quark
mass corresponds to the pion mass of $160$ MeV in the continuum limit \cite{Bazavov:2011nk},
which is slightly above the physical value. However, for the Polyakov loop this small difference
from the physical value plays no role.
The parameters of the lattice simulations including the lattice gauge coupling $\beta=10/g^2$
and the strange quark mass in lattice units are shown in  Table \ref{tab1} along with
the corresponding temperatures. The last column of the table shows the accumulated statistics
for each $\beta$ value in terms of molecular dynamics time units. 
The lattice spacing $a$ is determined
from the $r_1$ parameter defined in terms of the zero-temperature static potential as
\begin{equation}
\left.r^2 \frac{d V}{d r}\right|_{r=r_1}=1.0,
\end{equation}
and we use the value $r_1=0.3106$~fm \cite{Bazavov:2010hj}.
We use the  parametrization of the lattice spacing and the quark masses as functions of the gauge  coupling 
$\beta$ along the lines of constant physics that are
given in Ref. \cite{Bazavov:2011nk}. The $\beta$ dependent normalization constant $c$ that enters Eq. (\ref{Ldef})
was also taken from Ref. \cite{Bazavov:2011nk}.
Since we are interested in the low temperature behavior of the Polyakov loop, we also performed
additional calculations on $32^3 \times 8$ lattices for three values of the temperature,
$T=116$ MeV, $125$ MeV and $131$ MeV. The corresponding simulation parameters are also given in Table \ref{tab1}. 
Our calculations extend to temperatures as low as $116$ MeV, which is lower than in any previous lattice study.
Our numerical results for the Polyakov loop are 
shown in Fig. \ref{fig:lren}. 
To obtain continuum results for the renormalized Polyakov loop we first perform
a smooth spline interpolation of the numerical data for each $N_{\tau}$. 
The errors of the spline interpolation are determined using the bootstrap method.
Then we perform continuum extrapolations at selected temperature values
from $T=120$ MeV to $210$ separated by $5$ MeV steps. In addition we also consider
the renormalized Polyakov loop at $T=117$ MeV.
Since the leading discretization errors in the staggered 
fermion formulation are proportional to $a^2$,
we expect that for the renormalized Polyakov loop they should scale like $(a T)^2=1/N_{\tau}^2$.
Therefore we performed $1/N_{\tau}^2$ extrapolation of the renormalized Polyakov loop for
$T \ge 135$ MeV, where we have at least three lattice spacings. At lower temperatures
we have only two lattice spacings to estimate the continuum limit, corresponding to $N_{\tau}=8$ and $10$.
Furthermore, as one can see from Fig. \ref{fig:lren} (right) the lattice data do not show a clear $N_{\tau}$
dependence at these temperatures within the estimated errors. Moreover, the ordering of the $N_{\tau}=8$ and
$N_{\tau}=10$ data seems to be the opposite of that at $T>135$ MeV. For this reason we estimate
the continuum limit for $L_{ren}$ at $T \le 135$ MeV by averaging the interpolated $N_{\tau}=8$ and $N_{\tau}=10$ data.

Our continuum estimates for the renormalized Polyakov loop are also shown in Fig. \ref{fig:lren} and are
compared with the continuum results obtained using the stout action \cite{Borsanyi:2010bp}. 
The two lattice extrapolated continuum results agree with each other, except for $T=140$ MeV, 
where our results are larger by two standard deviations. 
\begin{table}
\begin{tabular}{|c|c|c|c|}
\hline
$ \beta $  & $m_s$  &  $T$ [MeV]  &  $\# TU$ \\
\hline
\multicolumn{4}{|c|}{$N_{\tau}=10$}          \\
\hline
6.285    & 0.0790 &  117        & 2423       \\ 
6.341    & 0.0740 &  123        & 7679       \\ 
6.390    & 0.0694 &  129        & 4990       \\ 
6.423    & 0.0670 &  133        & 3640       \\
6.460    & 0.0642 &  138        & 4200       \\ 
6.488    & 0.0620 &  142        & 3370       \\
6.515    & 0.0604 &  146        & 4988       \\ 
6.550    & 0.0582 &  151        & 4990       \\ 
6.575    & 0.0564 &  155        & 4990       \\ 
6.608    & 0.0542 &  160        & 4990       \\ 
6.664    & 0.0514 &  168        & 5000       \\ 
6.700    & 0.0496 &  174        & 4990       \\ 
6.740    & 0.0476 &  181        & 4990       \\ 
6.770    & 0.0460 &  186        & 4990       \\ 
6.800    & 0.0448 &  192        & 5310       \\
6.840    & 0.0430 &  199        & 4990       \\
6.880    & 0.0412 &  207        & 4990       \\
\hline
\multicolumn{4}{|c|}{$N_{\tau}=8$}           \\
\hline
6.050    & 0.1064 &  116        & 3977       \\
6.125    & 0.0966 &  125        & 3180       \\
6.175    & 0.0906 &  131        & 3732       \\
\hline
\end{tabular}
\caption{Simulation parameters for $40^3 \times 10$ 
and $32^3 \times 8$ lattices. The last column shows the accumulated statistics in terms of molecular dynamics trajectories.}
\label{tab1}
\end{table}   

As discussed in section I the deconfinement phase transition is related to $Z(N)$ symmetry 
in the case of infinitely heavy quarks. In the opposite limit of massless quarks there is
a chiral restoring phase transition. The connection between
the deconfinement crossover and the chiral crossover in QCD with the physical values of the
quark masses is a subject of long-standing discussions (see e.g. Refs. \cite{Digal:2000ar,Mocsy:2003qw,Hatta:2003bc}). 
Therefore, it is interesting to compare the temperature dependence of the renormalized Polyakov loop
in the continuum limit to the temperature dependence of the chiral condensate which is
used to described the chiral aspects of the QCD crossover. 
Combining our numerical results with the published HotQCD \cite{Bazavov:2011nk} results we estimated the renormalized
chiral condensates $\Delta_{ls}$ and $\Delta_l^R$ defined in Ref. \cite{Bazavov:2011nk} in
the continuum limit. So far continuum extrapolated data for the chiral condensate are only
available for the stout action.
The details of this analysis are given in the Appendix, where we also compare our results
with the one obtained using the stout action. 
We also calculated the strange quark condensate $\Delta_s^R$, which is analogous to $\Delta_l^R$ (see Ref. \cite{Bazavov:2011nk}) 
in the continuum limit. The details of these calculations are also given in the Appendix. 

In Fig. \ref{fig:lren_pbp}
the temperature dependence of the Polyakov loop is compared with the temperature dependence
of the renormalized chiral condensate as well as with the temperature dependence of the strange
quark condensate. As one can see from the figure the renormalized Polyakov loop changes very
smoothly in the temperature interval where the chiral condensates drops rapidly and it is
difficult to tell whether the transition in the renormalized Polyakov loop and chiral condensates
are connected. Comparison with the hadron gas model described in the next section, however, can
provide further insight into this issue.
Finally, the strange quark condensate shows a smooth behavior similar to that of $L_{ren}$.
\begin{figure}
\includegraphics[width=8.6cm]{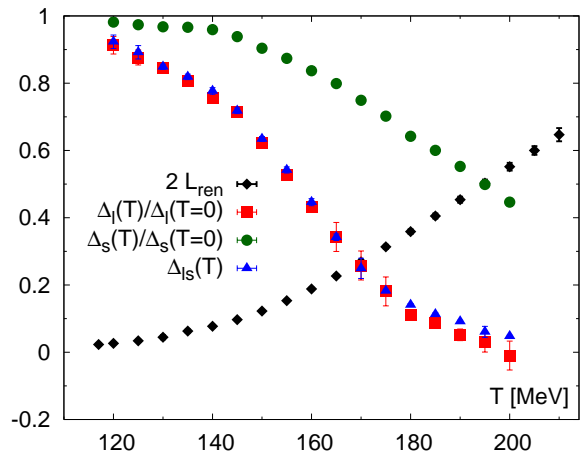}
\caption{The temperature dependence of the renormalized Polyakov loop compared
to the temperature dependence of the renormalized chiral condensates $\Delta_l^R$
and $\Delta_{ls}$ as well as the strange quark condensate $\Delta_s^R$. The values
of $\Delta_l^R$ and $\Delta_s^R$  have been normalized by the corresponding
zero temperature values. 
$\Delta_{ls}$ goes to one in the zero temperature limit by construction.
All results are continuum extrapolated.}
\label{fig:lren_pbp}
\end{figure}

\section{The hadron gas model}
As discussed in section I, at very low temperature the free energy of
a static quark is largely determined  by the binding energy of the lowest static-light meson.
In addition, there are contributions from static-strange mesons 
and baryons with one static quark.
Thus, following Ref. \cite{Megias:2012kb}, at very low temperature the Polyakov
loop is given by the contribution of the lowest static-light states
\begin{eqnarray}
&
3 L_{ren}=4 \exp(-M/T)+2 \exp(-M_s^0/T)+\nonumber\\
&
\sum_I \sum_j (2I+1) (2j+1) \exp(-M^{B0}_{I,j}/T),
\end{eqnarray}
where the spin and iso-spin degeneracies of the meson states have been taken into account and 
the summation of all iso-spin ($I$) states, as well as of 
all light and/or strange quark angular momentum states ($j$) for the lowest static-light baryons.
Altogether we have contributions from 6 mesons and 21 baryons \cite{Megias:2012kb}. 
The factor $3$ on the left-hand side of the above equation originates from
the color normalization in the definition of $L_{ren}$ in Eq. (\ref{Ldef}) and can
be seen from the spectral decomposition of the Polyakov loop correlator derived in
\cite{Jahn:2004qr}. Interactions with the medium are suppressed at  very
low temperatures but start to become more important as the temperature
increases. One may try to include the interaction by assuming that it can
be approximated by resonances. This assumption seems to work quite well
for bulk thermodynamic quantities 
\cite{Huovinen:2009yb,Borsanyi:2010bp,Borsanyi:2011sw,Bazavov:2012jq}.
The static-light meson contains a divergent self-energy contribution
which needs to be subtracted. This leaves the mass of the ground 
static meson state undetermined, while the masses of all other states
are given with respect to the mass of the lightest state $E_i=M_i-M_0$.
Therefore, we can generalize
the above equation as follows:
\begin{eqnarray}
&
L_{ren}=\frac{1}{3} \exp(-\Delta/T)(4 +2 \exp(-E_0^s/T)+\nonumber\\
&
\sum_{n,I,j} (2I+1) (2j+1) \exp(-E_{n,I,j}/T)).
\label{lren_hrg}  
\end{eqnarray} 
Here $E_0^s$ is the energy of the lightest static-strange meson with respect
to the mass of the lowest static-light state $M_0$.
The first and the second terms correspond to the contribution of the ground state static-light and
static-strange mesons, while the third term corresponds to the contribution of
the baryon states and the excited meson states. The index $n$ denotes different
excited states corresponding to the same values of $I$ and $j$. The renormalized
Polyakov loop depends on the subtracted mass of the lowest static-light meson
$\Delta$. This needs to be adjusted to match the lattice data for the Polyakov
loop at low temperatures, i.e. $\Delta$ should be adjusted to the specific scheme
used for the normalization of the Polyakov loop on the lattice. This matching procedure will
be discussed in subsection C.
In the following two subsections we are going to discuss
the meson and baryon contributions to $L_{ren}$ separately.

\subsection{Static mesons and their contribution to the renormalized Polyakov loop}

Static-light and static-strange mesons are characterized by the angular
momentum of the light (strange) quark and parity $j^P$. The spectrum of static-light
mesons consists of approximately degenerate pairs with $j=|l\pm 1/2|$, where
$l$ is the orbital angular momentum. 
The spectrum of static-light(strange) mesons has been studied in 2-flavor lattice QCD 
by the ETMC collaboration for $j$ up to $7/2$ that correspond to orbital angular
momentum $l=0,1,2$ and $3$ and are denoted by $S,~P_{\pm},~D_{\pm},F_{\pm}$ \cite{Michael:2010aa}.
Also the masses of first static-light and static-strange excited meson states for
$1/2^-$ channel (first radial or $S^*$ state in the ETMC notation ) have been calculated \cite{Michael:2010aa}.
To get rid of the divergent self-energy contribution, the masses of different states
are calculated with respect to the ground state ($S$ state) mass both in the light and
the strange quark sectors.  
These mass differences have been extrapolated to the continuum limit and 
to the physical pion mass and are approximately the same for static-light and static-strange
mesons, see Table 5 of Ref. \cite{Michael:2010aa}. The errors for the above mass
difference vary between $12$ MeV  and $37$ MeV. With all the spin and iso-spin degeneracies
the number of states identified on the lattice is $96$.
To calculate the Polyakov loop according to Eq. (\ref{lren_hrg}) we need to know $E_0^s$,
the energy (mass) of the lowest static-strange meson with respect to the lightest
static-light meson.We use phenomenological considerations to do so. Consider the
spin-averaged mass of the ground state charmed (bottom) mesons with strangeness $S=0$
and $S=-1$:
\begin{eqnarray}
&
\displaystyle
\overline{M}_D=\frac{3 M(D^*)+M(D)}{4}=1975~{\rm MeV},\\
&
\displaystyle
\overline{M}_B=\frac{3 M(B^*)+M(B)}{4}=5314~{\rm MeV},\\
&
\displaystyle
\overline{M}_{Ds}=\frac{3 M(D_s^*)+M(D_s)}{4} =2076~{\rm MeV},\\
&
\displaystyle
\overline{M}_{Bs}=\frac{3 M(B_s^*)+M(B_s)}{4}=5404~{\rm MeV}.
\end{eqnarray}
Here we used the values of the charm and bottom meson masses from Particle Data Group \cite{pdg}.
We get $\overline{M}_{Ds}-\overline{M}_D=100$ MeV and $\overline{M}_{Bs}-\overline{M}_B=90$ MeV. 
Heavy quark effective theory predicts that the masses of heavy-light mesons and thus also
the above difference should scale as the inverse of the heavy quark mass $m_Q$. Using this and 
the values of $\overline{M}_D$ and $\overline{M}_B$ as proxies for the charm and bottom quarks
respectively we get a value of $84$ MeV for the difference of the lowest static-strange
and static-light meson mass, i.e. $E_0^s=84$ MeV for $m_Q=\infty$. 
It is interesting to mention that the value
of the strange quark mass in the $\overline{MS}$ scheme, $m_s(\mu=2~{\rm GeV})=95(5)~{\rm MeV}$,
is close to the  value of $E_0^s$.
So $E_0^s$ may be interpreted as a constituent strange quark mass. 

To study the contribution of higher excited states we will use the $D_s$ meson spectrum
calculated on the lattice \cite{Bali:2011dc} as well as in a relativistic quark model 
\cite{Godfrey:1985xj,Godfrey:1986wj,Ebert:1997nk,Ebert:2011zz}.
The spectrum of $D_s$ mesons has been calculated on the lattice using improved  Wilson fermion actions 
\cite{Bali:2011dc}.
One needs to establish a relation between the meson masses in the static case and the masses of $D_s$ mesons.
Heavy mesons, $D_s$ mesons in particular, are characterized by $n L_J$, with $J$ being the total
angular momentum of the meson, $L$ being the orbital momentum and $n$ being the radial quantum number.
Obviously the $j=0$ states in the static limit are identified with spin-averaged $S$-state $D_s$ mesons. 
For a finite heavy quark mass, the $L-1/2$ state 
in the static limit splits into two states, $L_{L-1}$ and $L_L$, while the $L+1/2$ state 
splits into $L_L$ and $L_{L+1}$ state, e.g.,
$P_{-}$ becomes $1P_0$ and $1P_1$, and $P_{+}$ becomes $1P_1$ and $1P_2$. 
The corresponding splittings, however, are small.
Such a degeneracy pattern is indeed observed in the experimentally established positive 
parity ($P$-wave) $D$ and $D_s$ mesons.
Furthermore, the mass differences of various $D_s$ meson states calculated on the 
lattice and the spin-averaged lowest $S$-state are
in reasonable agreement with the mass differences in the static limit discussed above. 
Therefore, it is justified to use
the $D_s$ meson masses calculated on the lattice as proxies for static-strange mesons. 
Since the above difference is
approximately the same for strange and light quark cases we can use 
the same mass difference also for the light quarks.
That allows us to include the following excited states into the analysis: 
$2P$, $2D$, and $2F$. We identify the mass of
$2 L_{-}$ meson in the static case with the lowest $2L_{L}$ $D_s$ meson mass 
and the $2L_{+}$ with the higher  $2L_{L}$ $D_s$ meson.
With all the spin-iso-spin degeneracies this gives $90$ states.

To include even higher excited states we use quark model predictions.
The quark model can predict certain qualitative features of the heavy-light 
and static-light meson spectrums correctly  
\cite{Godfrey:1985xj,Godfrey:1986wj,Ebert:1997nk}.
However, the quark models also have problems. 
In the static limit the mass of the $P_{+}$ state is smaller than the mass of the $P_{-}$ state
just the opposite of what is observed on the lattice \cite{Michael:2010aa}. 
The mass of the $P_{+}$ in a quark model, 
calculated in Ref. \cite{Ebert:1997nk},
is $230$~MeV below the lattice result. Similarly, the mass of the $2S$ 
state is $441$ MeV below the lattice result \cite{Michael:2010aa}.
Comparing the results of Refs. \cite{Godfrey:1985xj,Godfrey:1986wj} to Ref. \cite{Ebert:1997nk},
one may conclude that the model dependence is small for the $1P$ meson
states, however, in the $B_s$ sector the masses of the $2S$ states differ by about $300$~MeV.
We try to account for these problems by assigning a theoretical error to the masses of higher
excited states.
 
\begin{figure}
\includegraphics[width=9cm]{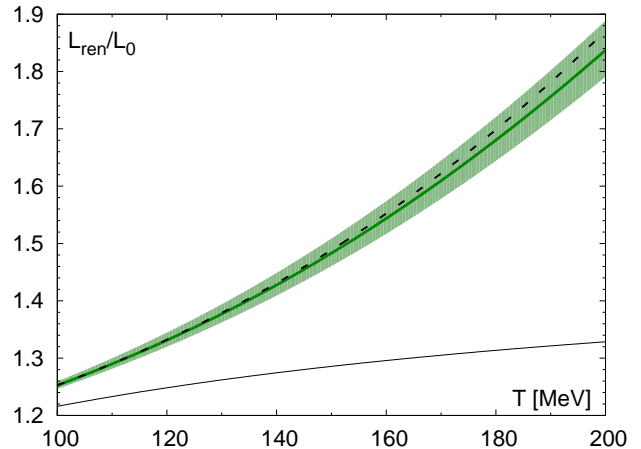}
\caption{Meson contribution to $L_{ren}$ normalized by $L_0=4 \exp(-\Delta/T)/3$. 
The band shows the contribution
of all static meson states calculated on the lattice (see text). 
The dashed line includes the contribution of higher
excited states. The thin solid line corresponds to the 
contribution of the lightest static-strange meson only.}
\label{fig:lren_mes}
\end{figure}

We took the results of quark model calculations of $D$ and $D_s$ mesons \cite{Ebert:2011zz} to 
estimate the contribution of $3S$, $4S$, $5S$, $3P$, and $1G$ states.
From comparison of the results of different quark models, as well as the comparison to the 
lattice results discussed above, we estimate the uncertainty of the masses
of the higher $nS$ states ($n\ge 3$) to be $300$~MeV, while for 
the other states we estimate it to be $150$~MeV.
It turns out, however, that contribution of these states to $L_{ren}$ is negligible up to temperatures
of $210$~MeV for which the model makes sense.
The excited states discussed so far should include all the possible states up to masses
of $2$ GeV above the ground state mass.
It is unlikely that individual resonance states can be observed above that energy.
In Fig. \ref{fig:lren_mes} we show the contribution of meson states to the renormalized Polyakov loop.
We normalize the results by $L_0=4 \exp(-\Delta/T)/3$. The contribution of all static meson states calculated
on the lattice in Ref. \cite{Michael:2010aa} is shown
as the band, the dashed line includes the contribution of the higher excited states, and the thin solid
line corresponds to the first static-strange meson only. The uncertainty band is determined by the errors
of the static meson masses calculated on the lattice \cite{Michael:2010aa}.
We see that at low temperatures the only meson state that
has a significant contribution to $L_{ren}$ in addition to the lowest state is the 
lightest static-strange meson, though the contribution of excited states cannot be completely neglected.
Excited states (up to $1F$) become very important at higher temperatures, $T>140$ MeV. Finally, 
the contribution of higher excited states is quite small and is only visible for $T>170$ MeV.

\subsection{Baryon contribution to the renormalized Polyakov loop}
The spectrum of baryons with one static quark has been studied by the ETMC collaboration \cite{Wagner:2011fs}
in 2 flavor QCD. The masses of static baryons with positive and negative parity and angular momentum
of light quarks $j=0$ and $1$ have been calculated. These states correspond to the ground state and first
orbital excitation of baryons with one heavy quark, i.e., to $1/2^+$, $3/2^+$ and $1/2^-$ and $3/2^-$. 
Counting the iso-spin and angular momentum degeneracies, the lowest positive and negative parity baryons
correspond to $79$ states.
The calculations have been carried out at one lattice spacing. The lack of continuum extrapolation
is not of great concern, since based on the studies of the static meson spectrum, 
cutoff effects are expected to be small compared to the statistical errors.  
The results presented in Ref. \cite{Wagner:2011fs} depend somewhat on how the lattice spacing is set. More
precisely, using $f_{\pi}$ the lattice spacing was determined to be $0.079(3)$~fm, while using the nucleon
mass the lattice spacing turned out to be $0.089(5)$~fm. 
In our analysis we use the values of the masses obtained by
fixing the lattice spacing through the nucleon mass $m_N$, since this procedure gives a value
of the $r_0$ parameter that is consistent with other determinations \cite{Aoki:2009sc,Bazavov:2011nk}, 
namely $r_0=0.473 \pm 0.09 (stat.) \pm 0.16(syst.)$ fm \cite{Alexandrou:2008tn}.
Setting the scale with $f_{\pi}$ gives $r_0=0.42$~fm \cite{Baron:2009wt} 
which is much smaller than any other determination. The lowest lying positive and
negative parity baryons give a fairly large contribution to
the renormalized Polyakov loop, in fact, the largest contribution next to the ground state mesons.
However, it turns out that higher excited states cannot be neglected.
We can use quark models to estimate the contribution of higher lying baryon states to $L_{ren}$.
 
The spectrum of baryons containing one heavy ($c$ or $b$) quark has been studied
in the relativistic quark model \cite{Capstick:1986bm} and in the relativistic quark-diquark 
model \cite{Ebert:2011kk}.
The analysis of Ref. \cite{Capstick:1986bm}, however, was restricted to $\Lambda_{b,c}$ 
and $\Sigma_{b,c}$ baryons.
We will use the spectrum of excited heavy baryons containing a $b$ quark as a proxy for 
the spectrum of higher excited baryon states with a static quark.
The masses of baryons with a static quark are  determined 
by the  angular momentum of the light quarks $j$. Therefore,
the heavy baryons form doublets with almost the same mass that correspond to the same angular momentum $j$
of the light quarks and total angular momentum $J=j \pm 1/2$. 
In our analysis we consider the mass difference of the baryon  
with the lower angular momentum in the doublet and the spin averaged mass of $B(B^*)$ mesons. These
mass differences obtained in a quark model are compared to the static baryon spectrum for
the lowest positive and negative parity states calculated on the lattice.
It turns out that the agreement between the lattice results and the model calculations is quite good
if the nucleon mass is used to set the lattice spacing. In fact, the lattice results agree with
the model calculations within the errors. For the $\Lambda_b$ and $\Sigma_b$ families we also find
good agreement between the diquark model and Ref. \cite{Capstick:1986bm} for the lowest states
of both parities.
In our calculations we use the spectrum calculated in Ref. \cite{Ebert:2011kk} which corresponds
to baryon states with angular momenta up to $J=11/2$ equivalently to $j=5$ of the light quarks and
up to 5 radial excitations. Counting all the spin and iso-spin degeneracies these correspond 
to $984$ states.

\begin{figure}
\includegraphics[width=9cm]{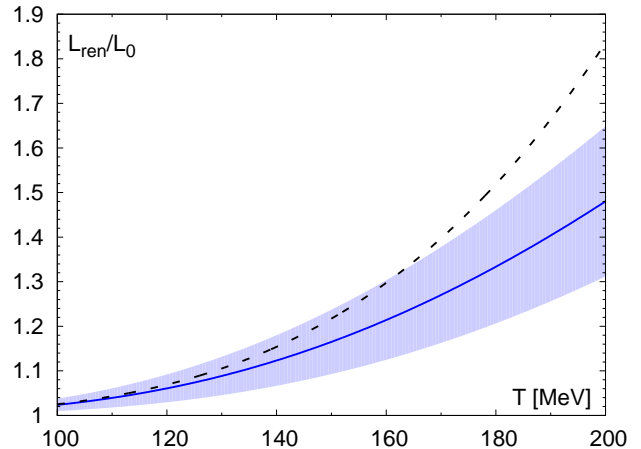}
\caption{The contribution of baryons to $L_{ren}$ normalized by $L_0=4 \exp(-\Delta/T)/3$. The solid line
and the band correspond to the contribution of the lowest positive and negative parity baryons, while
the dashed line corresponds to the contribution of all baryon states.}
\label{fig:lren_bar}
\end{figure}

As discussed above, different model calculations agree with each other for the lowest 
positive and negative parity states.
Unfortunately, the agreement is not that good for the higher excited states. To estimate the 
sensitivity of the Polyakov loop to the model
uncertainty of the higher excited baryon states we calculated the contribution of excited 
$\Lambda_Q$ and $\Sigma_Q$ baryons to $L_{ren}$, including all states up to
$J=7/2$, using the results of Ref. \cite{Capstick:1986bm} and of the diquark model \cite{Ebert:2011kk}.
The contributions of $\Lambda_b$ to $L_{ren}$ are a factor of two larger if one uses the spectrum from  
Ref. \cite{Capstick:1986bm} compared to the case where the $\Lambda_b$ 
spectrum from the diquark model is used. On the other hand, the contribution of the $\Sigma_b$ baryons
is a factor of two smaller if one uses the results of Ref. \cite{Capstick:1986bm} instead of the results
of the diquark model. Therefore we estimate that contribution of the higher excited baryon states 
is uncertain by a factor $2.5$. This is the largest source of uncertainty in the hadron resonance gas model 
for $T>170$~MeV.
The contribution of baryons to the renormalized Polyakov loop is shown in Fig. \ref{fig:lren_bar}.
The contribution of the static baryons identified on the lattice in Ref. \cite{Wagner:2011fs} is shown
as the solid line and the band. The error band corresponds to the uncertainty that has been evaluated using
the errors on the static baryon masses. The contribution of all baryon states to $L_{ren}$ is shown as
the dashed black line. At temperatures $T<120$ MeV the contribution of the baryons is below $10\%$. It becomes
significant above that temperature. The contribution of higher excited states becomes significant
only for $T>140$~MeV. Therefore, as  will become clear in the next subsection, the uncertainty 
in the comparison
to the lattice data due to the excited states is small.

\subsection{Comparison with lattice results}
Let us compare the hadron resonance gas model results with the lattice data discussed
in section II. The comparison is easiest in terms of the free energy of an isolated static
quark $F_Q(T)=-T \ln L_{ren}(T)$. 
The renormalization procedure of the Polyakov loop on the lattice introduces a scheme dependence.
Therefore, for the comparison of the hadron resonance gas with the lattice
data one needs to adjust the parameter $\Delta$
in Eq. (\ref{lren_hrg}).
We fix $\Delta$ by requiring that the hadron resonance gas model matches the continuum
lattice result at the lowest temperature $T=117$ MeV. This gives $\Delta=593 \pm 18$ MeV.
Once this constant is fixed, the hadron resonance gas model can predict the free energy of an isolated static
quark at any other temperature.

\begin{figure}
\includegraphics[width=9cm]{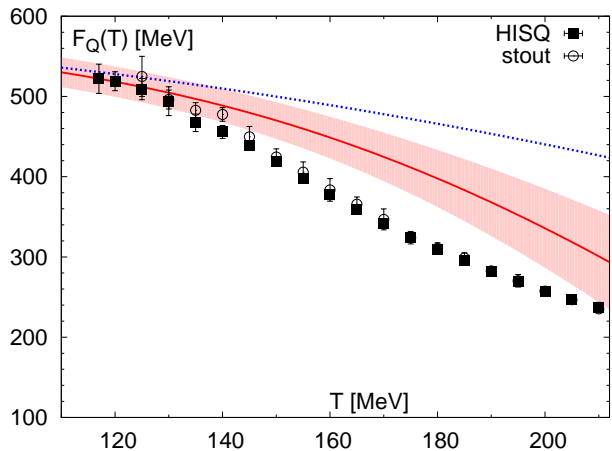}
\caption{The free energy of a static quark $F_Q(T)$ calculated on the lattice and compared
with the resonance gas model (solid line). The uncertainty of the hadron resonance gas model
is indicated by the band. The dotted line is the hadron resonance gas model result
with the ground state meson and baryon contribution only.}
\label{fig:FQ_comp}
\end{figure}

The comparison of the lattice data with the hadron resonance gas model is shown in Fig. \ref{fig:FQ_comp}.
The solid line and the band correspond to the hadron resonance gas results with all the states
discussed above and their uncertainty, which also includes the uncertainty in the value of $\Delta$.
The dashed line corresponds to the contribution of
the ground states only.
The figure shows that the contribution of excited states is significant already for
the lowest temperature available in the lattice calculations. 
The hadron resonance gas model can describe the lattice results on the renormalized Polyakov
loop up to temperature $140$ MeV, however, clearly fails above that temperature. 
All the excited states included in the analysis are not sufficient to explain the rapid
decrease of the static quark free energy. This result agrees qualitatively with the findings
of Ref. \cite{Megias:2012kb} if the stout data for $T<140$ MeV are used to normalize the static
free energy in their analysis (cf. Fig. 4 of Ref. \cite{Megias:2012kb}). 
More work is needed to understand the discrepancy between the lattice data and the hadron resonance
gas model. It is possible
that exotic hadron states can explain the discrepancy between the lattice results and the hadron
resonance gas model for $T>140$ MeV. Another possibility could be the partial restoration of
the chiral symmetry and the corresponding change in the static-light hadron masses, as indicated
in a recent lattice study \cite{Glozman:2012fj}.

\section{Conclusions}
We studied the renormalized Polyakov loop in lattice QCD using the HISQ action and obtained results in the continuum
limit for temperatures $120~{\rm MeV}<T<210~{\rm MeV}$. Results obtained with the HISQ action are in the $a^2$ scaling
regime for $N_{\tau}\ge 6$.
Our continuum results agree well with the earlier findings
obtained using the stout action \cite{Borsanyi:2010bp}. 
We also revisited the temperature dependence of the quark condensates and find that in the temperature region,
where the light quark condensates show a rapid decrease, the renormalized Polyakov loop changes very smoothly.
We do not see an obvious connection between the chiral and deconfinement transition described in terms of these
quantities.

We studied the question of the physics origin behind the increase
in the Polyakov loop, or equivalently, the decrease in the free energy of a static quark $F_Q$. At sufficiently high
temperatures the decrease in $F_Q$ is associated with the onset of color screening which also leads to the same decrease in
the  energy of a static quark at leading order \cite{Brambilla:2008cx}.
For temperatures $T<140$ MeV the decrease in $F_Q$ could be explained in terms of the hadron resonance
gas model. For larger temperatures the decrease in $F_Q$ appears to be significantly larger and
cannot be explained in terms of conventional static-light(strange) hadron states. It remains to
be seen whether this rapid decrease is due to the contribution from exotic static-light hadrons 
or some other mechanism. In the latter case its implication for color screening is not clear,
especially in view of recent
lattice results on the static energy of $Q \bar Q$ pair which
do not indicate large significant screening effects for $T<200$ MeV \cite{Bazavov:2012bq,Bazavov:2012fk}. 
Currently one of the largest uncertainties in the Polyakov loop calculations within the hadron resonance gas model 
comes from the excited baryon states. Clearly an improved lattice calculation of the static baryon spectrum  would 
be very helpful in this regard. Another open issue is the contribution of exotic static-light hadron states.

\section*{Appendix: Temperature dependence of the quark condensates}
\begin{figure}
\includegraphics[width=9cm]{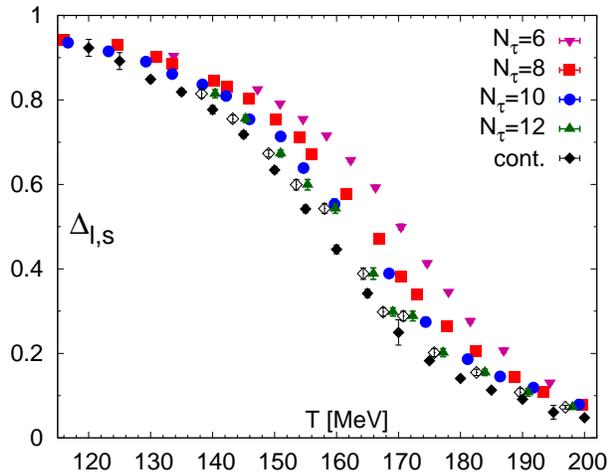}
\caption{The subtracted chiral condensate calculated with the HISQ action. The open diamonds
correspond to the $N_{\tau}=12$ results obtained using the $f_K$ scale \cite{Bazavov:2011nk} (see text).}
\label{fig:Deltals}
\end{figure}
In this appendix we discuss the calculation of the renormalized quark condensates.
The quark condensate $\langle \bar\psi \psi \rangle_q$ needs a multiplicative renormalization, and for non-zero
quark masses also an additive renormalization. It is easy to see that the leading
additive divergence is proportional to the quark mass and is quadratic in the cutoff
(inverse lattice spacing). Therefore, studying the following combination,
called the subtracted quark condensate, was proposed \cite{Cheng:2007jq}:
\begin{equation}
\Delta_{l,s}(T)=\frac{\langle \bar\psi \psi \rangle_{l,\tau}-\frac{m_l}{m_s} \langle \bar \psi \psi \rangle_{s,\tau}}
{\langle \bar \psi \psi \rangle_{l,0}-\frac{m_l}{m_s} \langle \bar \psi \psi \rangle_{s,0}}.
\end{equation}
Here $q=l$ and $s$ correspond to light and strange quarks, while the subscripts $x=0,\tau$ 
refer to zero and finite temperature expectation values, respectively. The expectation values 
$\langle \bar\psi \psi \rangle_{q,x}$ are normalized per single flavor. 
Sub-leading divergences proportional to the quark mass cubed and the logarithm of the cutoff are 
expected to be small for the physical
values of the light quark masses. We calculated $\Delta_{l,s}$ on $N_{\tau}=10$ lattices. 
Combining this with the published
results of the HotQCD collaboration obtained on $N_{\tau}=8$ and $12$ lattices \cite{Bazavov:2011nk}  
we performed a continuum extrapolation. 
First we interpolated the lattice data for each $N_{\tau}$ using a smooth spline and estimated the error of 
the spline by bootstrap analysis. 
Next we performed a continuum extrapolation at selected values of temperature separated by $5$ MeV 
within the interval $120~{\rm MeV} \le T \le 200$ MeV 
assuming a $1/N_{\tau}^2$ behavior. 
We studied the variation of the extrapolated result with varying the fit range in $N_{\tau}$. These
variations have been included in our final error estimate. 
For $T<170$ MeV the $N_{\tau}=6$ data have not been included in the analysis as they are incompatible with a $1/N_{\tau}^2$ behavior.
For $T\le 140$ MeV no $N_{\tau}=12$ data are available so the extrapolation 
had to rely on $N_{\tau}=8$ and $N_{\tau}=10$ data only. 
The numerical results for different $N_{\tau}$ and as the continuum extrapolations
are shown in Fig. \ref{fig:Deltals}. The continuum extrapolated results are slightly above the continuum results obtained
with the stout action \cite{Borsanyi:2010bp}. This difference is expected due to the slight difference in the 
light quark masses used in the two calculations,
namely, $m_l=m_s/20$ versus $m_l=m_s/27$ in Ref. \cite{Borsanyi:2010bp}. In Fig. \ref{fig:Deltals} 
we also show the $N_{\tau}=12$ HISQ data obtained
using the lattice spacing determined from the kaon decay constant $f_K$ \cite{Bazavov:2011nk}. 
These data are systematically above our continuum estimate.

Alternatively, we can get rid of the ultraviolet divergences in the quark condensate by considering 
the following combination, which
is called the renormalized quark condensate \cite{Bazavov:2011nk}:
\begin{equation}
\Delta_q^R=d+2 m_s r_1^4 ( \langle \bar \psi \psi \rangle_{q,\tau}-
\langle \bar \psi \psi \rangle_{q,0} ),~~~~q=l,s.
\label{DR}
\end{equation}
Here $d$ is a normalization constant that is related to the light quark condensate in the chiral limit.
More precisely, $d=2 m_s r_1^4 \langle \bar \psi \psi \rangle_{l,0}(m_l \rightarrow 0)$. With the values
of $m_s$ and $\langle \bar \psi \psi \rangle_{l,0}(m_l \rightarrow 0)$ from Ref. \cite{Bazavov:2009bb}
we get $d=0.0232244$. The quantity defined in Eq. (\ref{DR}) is closely related to the renormalized 
quark condensate $\langle \bar \psi \psi \rangle_R$
introduced in Ref. \cite{Aoki:2009sc}. Using our $N_{\tau}=10$ results and the published HotQCD 
results for $N_{\tau}=6,~8$, and $12$ we perform a continuum extrapolation for $\Delta_q^R$.
As for $\Delta_{l,s}$ we first perform a smooth spline interpolation and estimate the errors 
of the spline by bootstrap analysis. Then we perform a $1/N_{\tau}^2$ continuum extrapolation for selected values of the temperature separated 
by $5$ MeV in the interval $120 {\rm MeV} \le T \le 200$ MeV based on the interpolation and its errors.
We performed extrapolations using subsets of the available $N_{\tau}$ values and the differences in the obtained fit values for
$\Delta_q^R$ were treated as systematic errors and entered into our final error estimate.
For $T \le 140$ MeV the continuum extrapolations are based on $N_{\tau}=8$ and $10$ data only. 

\begin{figure*}
\includegraphics[width=8.5cm]{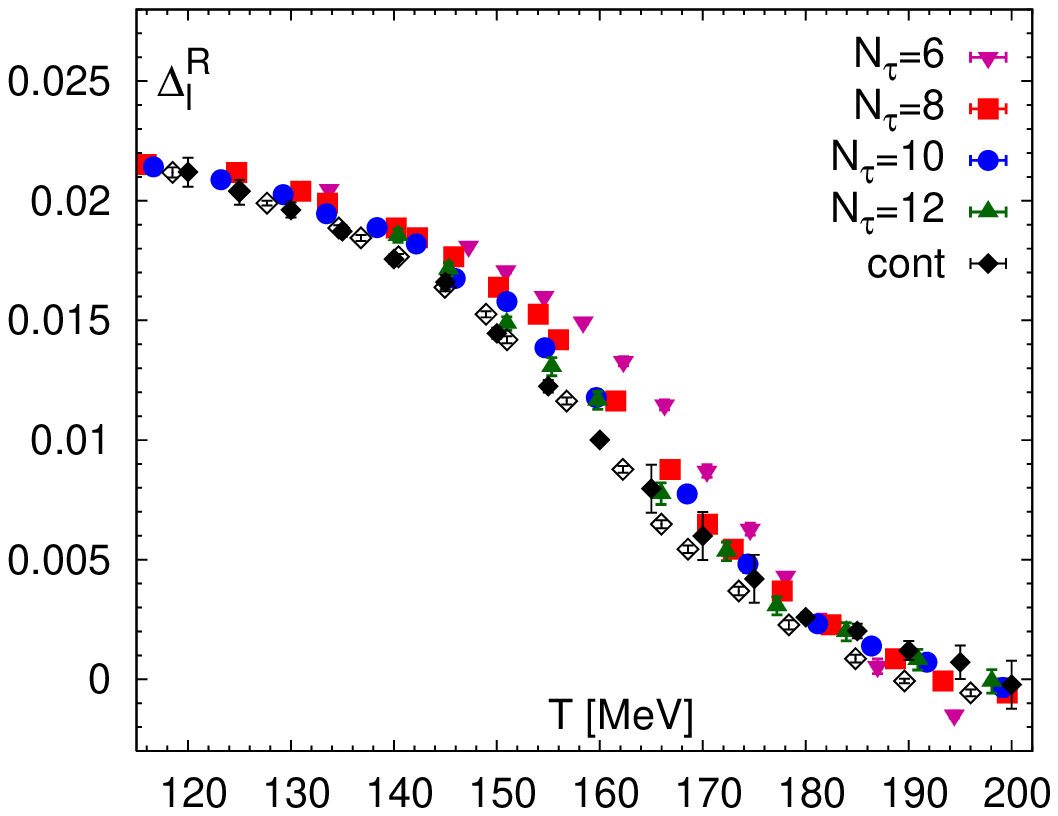}
\includegraphics[width=8.5cm]{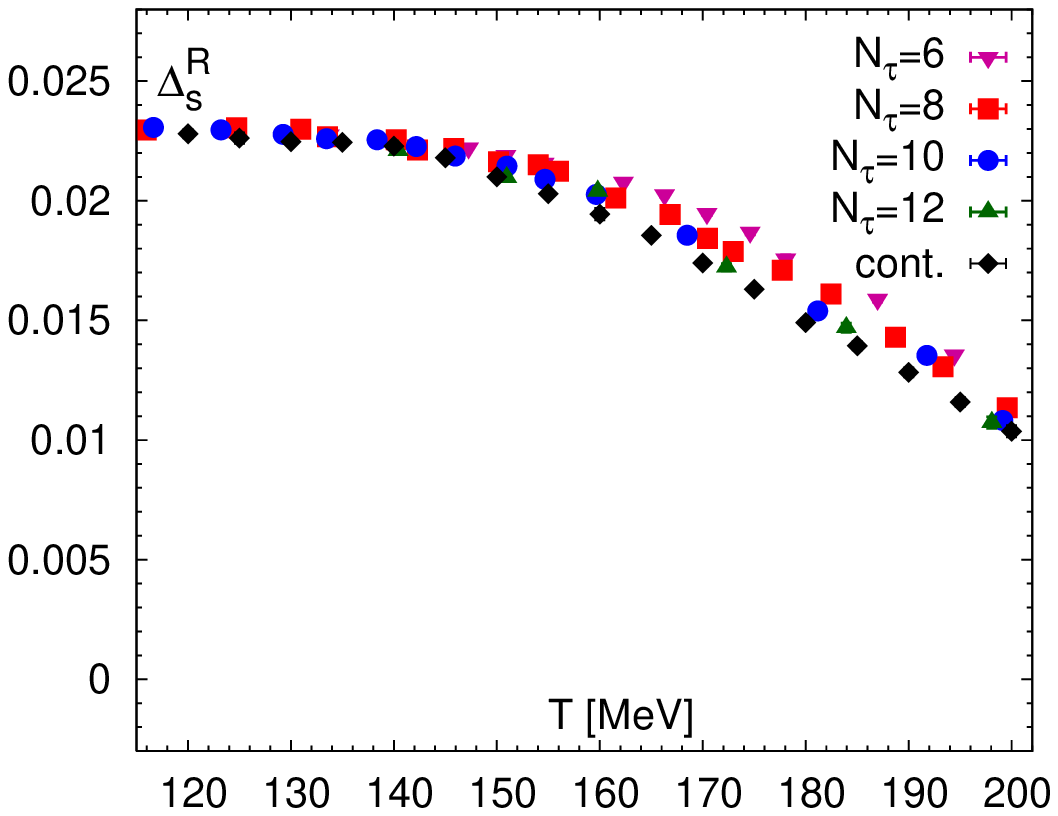}
\caption{The light (left) and strange (right) renormalized quark condensates calculated with the HISQ action. Open diamonds
correspond to $N_{\tau}=8$ HISQ results obtained with the $f_K$ scale \cite{Bazavov:2011nk}.}
\label{fig:pbpR}
\end{figure*}

The lattice QCD results for $\Delta_l^R$ and $\Delta_s^R$ are shown in Fig. \ref{fig:pbpR}, along with the continuum extrapolations. 
We also show
the HISQ $N_{\tau}=8$ data obtained using the lattice spacing from $f_K$ in this figure which seem to agree quite 
well with our continuum result,
except for $T>180$ MeV, where they are systematically lower.
Our continuum results for $\Delta_l^R$ are slightly larger than the continuum results obtained with the stout action \cite{Borsanyi:2010bp}. 
This is
again expected to be due to the difference in the light quark masses (see discussion in Ref. \cite{Bazavov:2011nk}).
Finally we would like to note the large difference in the temperature dependence of $\Delta_l^R$ and $\Delta_s^R$. 
The decrease of the renormalized
strange quark condensate is much more gradual than that of the light one and $\Delta_s^R$ reaches half of its vacuum value only at $T \simeq 200$ MeV.

\section*{Acknowledgments} 
This work was supported by the U.S. Department of Energy under
Contract No. DE-AC02-98CH10886. The numerical simulations
have been performed at NERSC and  on BlueGene/L computers at the New York Center for Computational
Sciences (NYCCS) at Brookhaven National Laboratory. The smooth spline interpolation and bootstrap
analysis was performed using the R package. We thank S. Mukherjee for his help with the R package and
F. Karsch for reading the manuscript and useful comments. We also thank E. Megias, 
E. Ruiz Arriola, and L.L. Salcedo for reading the first version of this paper and pointing out the normalization error
in the hadron resonance gas expression for the renormalized Polyakov loop, which we subsequently corrected.

\bibliographystyle{h-physrev.bst}
\bibliography{HotQCD}{}

\end{document}